\documentclass[conference]{IEEEtran}
\IEEEoverridecommandlockouts

\usepackage{cite}
\usepackage{amsmath,amssymb,amsfonts}
\usepackage{algorithmic}
\usepackage{graphicx}
\usepackage{textcomp}
\usepackage{comment}
\usepackage{xurl}
\usepackage{float}
\usepackage{subcaption}
\usepackage{enumitem}
\usepackage{makecell}
\usepackage{bm}
\usepackage[normalem]{ulem}
\usepackage{euscript}
\usepackage{mathtools}
\usepackage{siunitx}

\usepackage{titlesec}
\usepackage{xcolor}
\usepackage[thinlines]{easytable}

\usepackage{multirow}
\usepackage{fancyhdr}
\usepackage{booktabs}           
\usepackage{longtable}
\usepackage{tabularx}

\usepackage{hyperref}
\usepackage{array}

\renewcommand{\thesection}{\arabic{section}}
\renewcommand{\thesubsection}{\arabic{subsection}}
\renewcommand{\thesubsubsection}{\arabic{subsubsection}}

\colorlet{sectitlecolor}{red!60!black}

\colorlet{subsectitlecolor}{blue!60!black}

\colorlet{subsubsectitlecolor}{green!60!black}

\newlength{\anIndent}
\titleformat{\section}
  {\normalfont\Large\bfseries\color{sectitlecolor}}
  {\thesection)}{2\anIndent-\widthof{\thesection}}{}

\titleformat{\subsection}
  {\normalfont\large\bfseries\color{subsectitlecolor}}
  {\thesection.\thesubsection.}{2\anIndent-\widthof{\thesubsection}}{} 

\titleformat{\subsubsection}
  {\normalfont\normalsize\bfseries\color{subsubsectitlecolor}}{\thesection.\thesubsection.\thesubsubsection.}{1.5\anIndent-\widthof{\thesubsubsection}}{}

\newcolumntype{P}[1]{>{\centering\arraybackslash}p{#1}}

\usepackage{multirow}
\usepackage{fancyhdr}
\usepackage{booktabs}           
\usepackage{longtable}

\hypersetup{colorlinks, linkcolor=blue}

\def\BibTeX{{\rm B\kern-.05em{\sc i\kern-.025em b}\kern-.08em
    T\kern-.1667em\lower.7ex\hbox{E}\kern-.125emX}}
    
\renewcommand{\baselinestretch}{1.2}          

\DeclarePairedDelimiterX{\infdivx}[2]{(}{)}{#1\;\delimsize\|\;#2}
\newcommand{\infdiv}{\infdivx}

\makeatletter
\newcommand*\titleheader[1]{\gdef\@titleheader{#1}}

\AtBeginDocument{%
  \let\st@red@title\@title%
  \def\@title{%
    \bgroup\normalfont\large\centering\@titleheader\par\egroup
    \vskip1.5em\st@red@title}
}
\makeatother

\title{Synthetic ECG Signal Generation using Probabilistic Diffusion Models\\

\thanks{*This work was partially supported by Open Cloud Institute (OCI) at UTSA.}
}

\titleheader{Pre-print}

\begin{document}

\author{\IEEEauthorblockN{Edmond Adib*}
\IEEEauthorblockA{\textit{\footnotesize Electrical and Computer Engineering Department} \\
\textit{\small University of Texas at San Antonio}\\
\small San Antonio, USA \\
edmond.adib@utsa.edu}

\and

\IEEEauthorblockN{Amanda Fernandez}
\IEEEauthorblockA{\textit{\footnotesize Computer Science Department} \\
\textit{\small University Of Texas at San Antonio}\\
\small San Antonio,USA \\
amanda.fernandez@utsa.edu}

\and

\IEEEauthorblockN{Fatemeh Afghah}
\IEEEauthorblockA{\textit{\footnotesize Electrical and Computer Engineering Department} \\
\textit{\small Clemson University}\\
\small Clemson,USA \\
fafghah@clemson.edu}

\and

\IEEEauthorblockN{John J. Prevost}
\IEEEauthorblockA{\textit{\footnotesize Electrical and Computer Engineering Department} \\
\textit{\small University of Texas at San Antonio}\\
\small San Antonio, USA \\
jeff.prevost@utsa.edu}

}

\maketitle

\begin{abstract}
Deep learning image processing models have had remarkable success in recent years in generating high quality images. Particularly, the Improved Denoising Diffusion Probabilistic Models (DDPM) have shown superiority in image quality to the state-of-the-art generative models, which motivated us to investigate their capability in the generation of the synthetic electrocardiogram (ECG) signals. In this work, synthetic ECG signals are generated by the Improved DDPM and by the Wasserstein GAN with Gradient Penalty (WGAN-GP) models and then compared. To this end, we devise a pipeline to utilize DDPM in its original $2D$ form. First, the $1D$ ECG time series data are embedded into the $2D$ space, for which we employed the Gramian Angular Summation/Difference Fields (GASF/GADF) as well as Markov Transition Fields (MTF) to generate three $2D$ matrices from each ECG time series, which when put together, form a $3$-channel $2D$ datum. Then $2D$ DDPM is used to generate $2D$ $3$-channel synthetic ECG images. The $1D$ ECG signals are created by de-embedding the $2D$ generated image files back into the $1D$ space. This work focuses on unconditional models and the generation of \emph{Normal Sinus Beat} ECG signals exclusively, where the Normal Sinus Beat class from the  MIT-BIH Arrhythmia dataset is used in the training phase. The \emph{quality}, \emph{distribution}, and the \emph{authenticity} of the generated ECG signals by each model are quantitatively evaluated and compared. Our results show that in the proposed pipeline and in the particular setting of this paper, the WGAN-GP model is consistently superior to DDPM in all the considered metrics.
\end{abstract}


\section{Introduction}
 Electrocardiogram (ECG) is the manifestation of the heart's rhythm and electrical activity. ECG monitoring is a painless, fast, non-invasive, and cheap procedure which can reveal much regarding the heart's health. Automated ECG-based diagnosis is becoming increasingly more popular as it eliminates randomized human error and can be readily available at a patient's bedside using affordable wearable heart monitoring devices. Automatic ECG diagnosis models are usually deep learning models that classify patients' ECG signals according to the morphological pattern of the ECG \cite{saadatnejad2019lstm}. The datasets used for training these models are usually highly imbalanced, as normal beats are much more abundant than abnormal cases. Realistic synthetic\footnote{In this context \emph{synthetic beats} and \emph{generated beats} are the outputs of the generative models and used interchangeably.} ECG signals can augment real datasets, enriching them and enabling better class balance. Additionally, ECG data are considered private information, therefore their usage is highly regulated, whereas synthetic ECG signals can be used without any restriction.

In general, there are several approaches to addressing imbalances in datasets. One approach is to use oversampling methods such as SMOTE \cite{chawla2002smote} to generate additional synthetic data. Other approaches to address the inadequacy of samples are to utilize new designs of loss functions such as focal loss \cite{lin2017focal} or using a new training scheme, e.g., few-shot training \cite{wang2020generalizing}. However, using deep generative algorithms such as Generative Adversarial Networks (GAN) \cite{goodfellow2014generative} and Variational Auto-Encoders (VAE) \cite{doersch2016tutorial} is becoming more popular (e.g. \cite{frid2018synthetic} and \cite{kuznetsov2020electrocardiogram}).  

Computer vision deep learning models have been outstandingly successful in \emph{classification} as well as in \emph{generation} tasks. Most prominent models (AlexNet, VGG-16, ResNet-18, \dots) are pre-trained and readily available on the internet. Using transfer learning techniques, they need only a minor fine tuning to produce the best-quality results, saving days, weeks, or even months of training time. Additionally, using $2D$ data space provides more augmentation techniques (such as flipping, rotation, and mirroring), which are very beneficial particularly in the classification tasks. Moreover, better classification performances compared to $1D$ space implementations have been reported \cite{jun2018ecg}.
Improved Denoising Diffusion Probabilistic Models (DDPM) \cite{nichol2021improved} have proved to generate images with quality superior to GAN models \cite{dhariwal2021diffusion}. In this study, we present a pipeline to generate $1D$ synthetic ECG signals using $2D$ DDPM. We also investigate the quality of the generated beats by DDPM and compare it with GAN models. To the best of our knowledge, this is the first time that diffusion models are used for the generation of synthetic ECG signals. The $1D$ ECG time series data are embedded into $3$-channel $2D$ data (similar to RGB image files) using Gramian Angular Summation/Difference Fields (GASF/GADF) and Markov Transition Fields (MTF) \cite{wang2015imaging} and then fed to the Improved DDPM as image files. DDPM is trained on the embedded data and then sampled to generate embedded $2D$ ECG data which are then de-embedded and transformed back to $1D$ ECG time series using the inverse transformations (Figure. \ref{fig: pipeline}). Three different settings of DDPM hyper-parameters have been considered as three study cases and the fourth study case is the data generated by the WGAN-GP model. The four study cases are compared in terms of the \emph{quality}, \emph{distribution}, and \emph{authenticity} of the generated beats (such as whether they can replace the real data in a classification test). Precision score, Area Under Curve (AUC) of Precision-Recall curves, and also that of the Receiver Operating Characteristic curves (ROC AUC) have been used as the metrics of classification. MIT-BIH Arrhythmia dataset\cite{moody2001impact} \cite{goldberger2000components} is used as the dataset. Since this research is the first use of DDPM in ECG generation, we employed DDPM in the unconditional form and focused only on the \emph{Normal Sinus} class to investigate the feasibility of the idea. However, it can be also used in the conditional form for generating synthetic arrhythmia signals in various classes.

\subsection{Related Works}
Much research has been done on the generation of ECG in one-dimensional space. Alcaraz \emph{et al.} \cite{alcaraz2023diffusion} used a combination of structured state space model and diffusion models to generate 10-second 12-lead synthetic ECGs. They generated synthetic digital twins of PTB-XL ECG dataset. Wang \emph{et al.} \cite{wang2019ecg} augmented their imbalanced dataset using synthetically generated beats by a $1D$ auxiliary classier generative adversarial network (AC-GAN). Delaney \emph{et al.} \cite{delaney1909synthesis} studied the generation of realistic synthetic signals using a range of architectures from the GAN family. Hyland \emph{et al.} \cite{esteban2017real} employed two-layer BiLSTM architecture in the generator and the discriminator to generate synthetic ECG signals. Adib \emph{et al.} \cite{adib2021synthetic} compared several GAN models (mono-class) in one study, and in another study they compared conditional and unconditional GAN-GP models (multiclass) in the generation of synthetic ECG beats \cite{adib2022arrhythmia}. 

There are also research studies in which the $1D$ ECG time series are embedded into $2D$ space. Ahmad \emph{et al.} \cite{ahmad2021ecg} developed a novel image fusion model (by AlexNet architecture) for the classification task of ECG beats. They convert $1D$ beats into a $3$-channel $2D$ image using Gramian Angular Field (GAF), Recurrence Plot (RP), and Markov Transition Field (MTF). Cai \emph{et al.} \cite{cai2022electrocardiogram} also created a $3$-channel embedding using Gramian angular field (GAF), recurrent plot (RP), and tiling for their classification task. Diker \emph{et al.} \cite{diker2019novel} did $2D$ embedding of heartbeats using spectrograms to classify of them by AlexNet, VGG-16, and ResNet-18. Izci \emph{et al.} \cite{izci2019cardiac} mimicked LeNet CNN architecture for the classification of $2D$ grayscale images (plots) of beats without any embedding. Hao \emph{et al.} \cite{hao2019spectro} converted one-dimensional ECG signal into spectro-temporal images and used multiple dense convolutional neural networks to capture both beat-to-beat and single-beat information for analysis. Hao \emph{et al.} \cite{hao2019spectro} used a time-frequency representation of ECG beats (wavelet transform and short-term Fourier transform) for classification by a dense CNN model. Huang \emph{et al.} \cite{huang2019ecg} used a short-time Fourier transform to embed heartbeats into $2D$ spectrograms for their classification by a $2D$-CNN model. Oliviera \emph{et al.} \cite{oliveira2019novel} used wavelet transform for $2D$ embedding of beats. Salem \emph{et al.} \cite{salem2018ecg} employed spectrograms to classify ECG data as well. Mathunjwa \emph{et al.} \cite{mathunjwa2021ecg} and \cite{mathunjwa2022ecg} as well as \cite{zhang2021recurrence} used recurrent plots for classification of $2D$ ECG data. To the best of our knowledge, currently there is no study on using $2D$ models to \emph{generate} synthetic ECG signals.


\section{Methodology}
\subsection{Dataset and Segmentation}
 The MIT-BIH Arrhythmia dataset\cite{moody2001impact} \cite{goldberger2000components} is one of the benchmark datasets of ECG analysis. It is a set of $48$ Holter recordings, each $30$ minutes long and with two channels, that were obtained by the Beth Israel Hospital Arrhythmia Laboratory between 1975 and 1979. The upper channel is consistently the modified limb lead II (MLII) and the lower one is mostly a modified lead V1 (occasionally V2 or V5, and in one instance V4). The digitization of the analog signals has been done at $360$ Hz. In this study, we used the upper channel as it is only from one electrode (MLII).
 
 The \emph{Adaptive Window} method is used for the segmentation of the ECG signals into individual beats. This method adapts to changes in the heart rate. The R-peaks of the QRS complex in heartbeats are annotated in the dataset. Using this information, the distances to the next and previous R-peaks are determined for each individual beat and then $75\%$ of each is used as the \emph{cutoff} to find the boundaries of the individual beat. Then, all the segmented individual beats are resampled to $256$. The resulting dataset is comprised of $109,338$ individual beats in $15$ classes and is highly imbalanced.

\subsection{Denoised Diffusion Probabilistic Models}
\label{sec: DDPM Theory}

DDPM, similar to the VAE model, is a variational-based model \cite{nichol2021improved} where the objective is to find the distribution of the dataset explicitly \cite{kingma2019introduction}. There are two processes in the DDPM: forward process and backward process. In the forward process, $q$, noise is added to a datapoint $x_0 \sim q(x_0)$ gradually and in steps. The generated noisy samples (latent variables) are $x_1$ through $x_T$. The noise which is added at each time step $t$ is a Gaussian noise with a variance with a specific schedule $\beta_t$ \cite{ho2020denoising}:
\begin{equation}
    q(x_t|x_{t-1}) = \mathcal{N} \left(x_t; \sqrt{1-\beta_t}x_t, \beta_t \boldsymbol{I} \right)     
\end{equation}
 $q\left( x_t|x_0 \right)$ can be expressed in a closed form by defining $\alpha_t = 1-\beta_t$ and 
  $\Bar{\alpha}_t = \prod_{s=0}^{t}{\alpha_s}$:
\begin{equation}
    \begin{split}
        q(x_t|x_{0}) & = \mathcal{N} \left( x_t; \sqrt{\Bar{\alpha}_t}\,x_0, \left( 1-\Bar{\alpha}_t \right) \boldsymbol{I} \right) \\
         & = \sqrt{\Bar{\alpha}_t}\,x_0 + \epsilon \, \sqrt{1-\Bar{\alpha}_t}, \quad \epsilon \sim \mathcal{N} (0, \boldsymbol{I})
         \label{eq: DDPM 2}
    \end{split}    
\end{equation}
where, $1-\Bar{\alpha}_t$ is the variance of the noise for any arbitrary step and can be used directly instead of $\beta_t$. Using Bayes theorem, it can be proved that the posterior of the reverse process $q\left(x_{t-1}|x_{t}, x_0\right)$ is also a Gaussian distribution \cite{ho2020denoising}:
\begin{equation}
    q(x_{t-1}|x_{t}, x_0) = \mathcal{N} \left( x_{t-1}; \Tilde{\mu_t} \left( x_t, x_0 \right), \Tilde{\beta_t}\boldsymbol{I} \right)
\end{equation}
with:
\begin{equation}  
    \begin{cases}
      \Tilde{\mu_t}(x_t, x_0) & =  \frac{\sqrt{\Bar{\alpha}_{t-1}}\,\beta_t}{1-\Bar{\alpha}} \, x_0 + \frac{\sqrt{\alpha_t} \, (1-\Bar{\alpha}_{t-1})}{1-\Bar{\alpha}_{t}}    \\
        \Tilde{\beta_t} & = \frac{1-\Bar{\alpha}_{t-1}}{1-\Bar{\alpha_t}} \, \beta_t   \\        
    \end{cases}    
\end{equation}
To sample from $q(x_0)$, knowing the denoising process $q(x_{t-1}|x_{t})$, we can start from $q(x_T)$ and then sample the  reverse steps $q(x_{t-1}|x_{t})$ till we reach $x_0$. Under certain assumptions on $\beta_t$ and $T$, sampling $x_T$ is trivial, so pure noise is usually used \cite{ho2020denoising}. However, since $q(x_{t-1}|x_{t})$ is \emph{intractable}, a neural network is trained to approximate it. Since as $T\rightarrow\infty$, $\beta_t\rightarrow 0$ and $q(x_{t-1}|x_{t})$ approaches a diagonal Gaussian distribution, it would be sufficient to train a neural network to predict the mean $\mu_\theta$ and the diagonal covariance $\Sigma_\theta$ :
\begin{equation}
    p_\theta \left( x_{t-1}|x_{t} \right) = \mathcal{N} \left( x_{t-1}; \mu_\theta(x_t, t), \Sigma_\theta(x_t, t) \right)    
\end{equation}
The variational lower-bound loss function of $L_{vlb}$ for $p_\theta(x_0)$ is:
\begin{align}
    L_{vlb} & = L_0 + L_1 + \dots + L_{T-1} + L_T          \\
    L_0 & = -\log{p_\theta (x_0 \vert x_1)}             \\
    L_{t-1} & = D_{KL}\infdiv{q(x_{t-1} \vert x_t, x_0)}  {p_\theta(x_{t-1} \vert x_t)}           \\
    L_T & = D_{KL}\infdiv{q(x_{T} \vert x_0)}{p(x_T)} 
\end{align}
Ho \emph{et al.} \cite{ho2020denoising} used a simplified loss function in which a neural network $\epsilon_\theta(x_t, t)$ is trained to to predict $\epsilon$ from Eq. \ref{eq: DDPM 2}:
\begin{flalign}
     L_{simple} \hspace{-1mm}= \hspace{-1mm} 
       E_{t \sim [1,T],x_0\sim q(x_0),\epsilon \sim \mathcal{N}(0,\boldsymbol{I})} 
       \hspace{-1mm}\left[ \Vert\epsilon - {\epsilon_\theta(x_t, t)\Vert}^2 \right]\hspace{-1mm}       
    \label{eq: L Simple}
\end{flalign}
Then $\mu_\theta(x_t,t)$ can be derived from $\epsilon_\theta(x_t, t)$:
\begin{equation}
    \mu_\theta(x_t,t) = \frac{1}{\sqrt{\alpha_t}} \left( x_t - \frac{1-\alpha_t}{\sqrt{1-\Bar{\alpha}_t}} \, \epsilon_\theta(x_t, t) \right)
    \label{eq: appA 11}
\end{equation}
$L_{simple}$ does not provide any signal for training $\Sigma_\theta(x_t, t)$. So, instead of learning $\Sigma_\theta(x_t, t)$, it is fixed to a constant, i.e.,:
$\Sigma_\theta \left( x_t, t \right) = \sigma_t^2\boldsymbol{I}$ where $\sigma_t^2 = \beta_t$ or 
$\sigma_t^2 = \Tilde{\beta}_t = \frac{1-\Bar{\alpha}_{t-1}}{1-\Bar{\alpha_{t}}} \beta_t$ which are the upper and lower bounds for the true reverse process' step variance, respectively.

\subsection{Improved DDPM}
Log-likelihood is a metric in generative modeling and is an indication of the coverage of the existing modes in the dataset by the model \cite{razavi2019generating}. Also, it has been shown that it has a great impact on the quality of samples and learned feature representations \cite{henighan2020scaling}. Nichol \emph{et al.} \cite{nichol2021improved} made a few modifications on the DDPM developed by Ho \emph{et al.} \cite{ho2020denoising} to achieve better Log-likelihood, which are introduced below very briefly.

\subsubsection{Learned Sigma}
$\Sigma_\theta(x_t, t)$ is the variance of the reverse process. Nichol \emph{et al.} \cite{nichol2021improved} argue that by increasing the number of the diffusion steps, the role of the mean $\mu_\theta(x_t, t)$ becomes more dominant than $\Sigma_\theta(x_t, t)$ in determining the \emph{distribution} of the data. They also argue that not fixing, rather learning the $\Sigma_\theta(x_t, t)$ would provide a better choice  and the Log-likelihood is improved. They suggested an interpolation between $\beta$ and $\Tilde{\beta}$ (the two extremes) for the parameterization of the variance. In fact, their model outputs a vector $v$ which interpolates between the two fixed extreme values (lower and higher bounds) and has the same dimension as the data:
\begin{equation}
    \Sigma_\theta(x_t, t)  = \exp{\left(v\log \beta_t + \left(1-v\right) \log \Tilde{\beta}_t\right)}   
\end{equation}
Since $L_{simple}$ (Eq. \ref{eq: L Simple}) is independent of $\Sigma_\theta(x_t, t)$, they suggested a new hybrid objective to learn $\Sigma_\theta(x_t, t)$:
\begin{equation}
    L_{hybrid} = L_{simple} + \lambda \, L_{vlb}
\end{equation}
\noindent
in which $\lambda$ is set to a low value of $0.001$ to prevent $L_{vlb}$ from overwhelming $L_{simple}$.

\subsubsection{Noise Schedule}
Linear noising schedule is used in the original DDPM by Ho \emph{et al.} \cite{ho2020denoising}, which Nichol \emph{et al.} \cite{nichol2021improved} found sub-optimal for lower resolution image processing (i.e. $64\times64$ or $32\times32$). Instead, they proposed a \emph{cosine} schedule, which retain information in the noisy images longer in the noising process steps, as opposed to the strong linear schedule in which the noisy images become pure noise much earlier and destroys information more quickly in the noising process (Figures $3$ and $5$ in \cite{nichol2021improved}):
\begin{equation}
    \Bar{\alpha_t} = \frac{f\left(t\right)}{f\left(0\right)} \quad , \quad f\left(t\right) = \cos{{\left(\frac{t/T + s}{1 + s}\cdot\frac{\pi}{2}\right)}^2}
\end{equation}
\noindent
with $s$ being an offset parameter set at $s=0.008$.
\subsubsection{Importance-sampled \texorpdfstring{$L_{vlb}$}{Lvlb}}
Nichol \emph{et al.} \cite{nichol2021improved} found that, contrary to their expectation, optimizing $L_{hybrid}$ achieves better log-likelihood rather than optimizing $L_{vlb}$ directly, which they believe was caused by $L_{vlb}$ and its gradient being more \emph{noisy}. So, in order to reduce the variance of $L_{vlb}$, they employed \emph{importance sampling} rather than sampling uniformly throughout the samples:
\begin{equation}
    L_{vlb} \hspace{-1mm}=\hspace{-1mm}  \mathop{{}\mathbb{E}_{t\sim p_t}} \left[ \frac{L_t}{p_t}\right] \text{\footnotesize, where} \;  p_t \hspace{-1mm} \propto \hspace{-1mm} \sqrt{\mathop{{}\mathbb{E}}\left[ L_t^2 \right]} \;  \text{\footnotesize and}  \sum{p_t} = 1
\end{equation}
in which they kept a $10$-step history for the evaluation of $\mathop{{}\mathbb{E}}\left[ L_t^2 \right]$ which is updated dynamically. Using importance sampling, they could achieve their best log-likelihoods with considerably less noisy objective than the uniformly sampled objective. However, the importance sampling technique does not make any improvement on the less noisy $L_{hybrid}$ objective \cite{nichol2021improved}.  

\subsection{Pipeline}
\begin{figure*}  
    \begin{center}
        \includegraphics[scale=0.53, trim=10mm 70mm 5mm 50mm,clip]{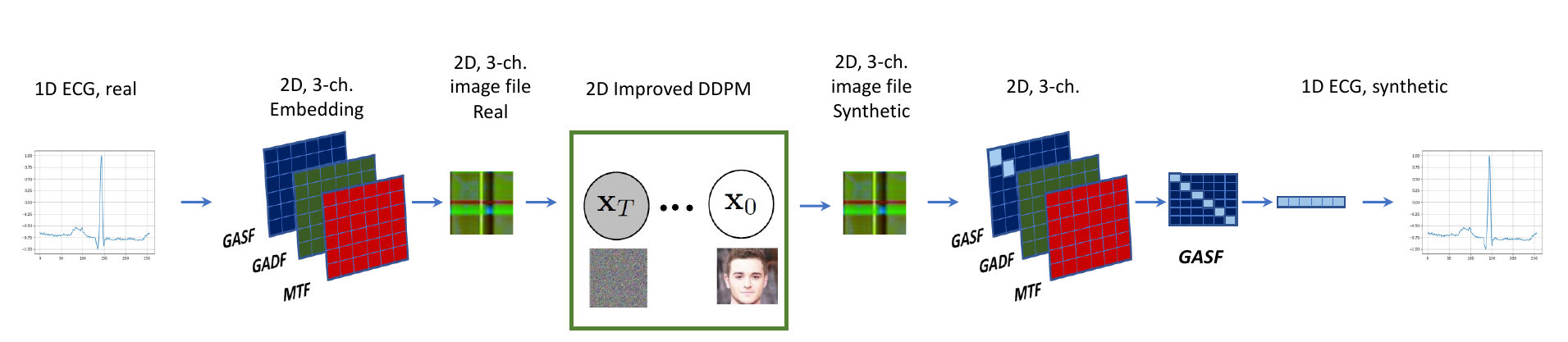}
        \caption{Block diagram of the proposed model. First, the normalized ECG time series are transformed into $2D$ space using GASF, GADF, and MTF separately, and $3$-channel RGB-like data are generated. Then, the Improved DDPM model is trained and sampled to generate $2D$ ECG data. Finally, the $1D$ ECG time series are reconstructed using the diagonals of the GASF channel.}
        \label{fig: pipeline}
    \end{center}
\end{figure*}

The general pipeline used in this study for the generation of synthetic ECG signals is shown in Fig. \ref{fig: pipeline}. First, $1D$ real ECG beats (exclusively in class \emph{N}) are embedded into the $2D$ space and then $3$-channel $2D$ image files (similar to RGB image files) are formed. Then, the Improved DDPM \cite{nichol2021improved} is trained and then sampled to generate $2D$ $3$-channel synthetic ECG image files. Finally, the $1D$ time series are reconstructed by de-embedding the generated data back into $1D$ space. The stages of the pipeline are discussed in more detail in the following sections. 

\subsection{1D-2D Embedding}
Wang \emph{et al.} \cite{wang2015imaging} proposed a novel embedding framework for mapping time series data from one-dimensional space into two-dimensional space, which enabled the utilization of computer vision techniques \emph{as-is} for time series analysis. First, they map the ECG time series from Cartesian to polar coordinates. Then, they use \emph{Gramian Angular Summation/Difference Fields} (GASF/GADF) and \emph{Markov Transition Fields} (MTF) to build $3$ separate $2D$ matrix embeddings of the time series and finally put them together to create a $3$-channel $2D$ image file similar to an RGB image file. The proposed pipeline in this study (Fig. \ref{fig: pipeline}) has been used only for the DDPM, whereas in the WGAN-GP model, all the data (training and generated) and the model itself are in the $1D$ space and no embedding was necessary.

\subsubsection{Polar Coordinates Representation}
ECG time series are one-dimensional, vector-like data, $X = \{x_1, x_2, \ldots, x_N\}$, which represent the time-progression of the induced voltage to the electrode caused by the motion of the electrical impulse generated by the sinoatrial (SA) node in the heart. When normalized and rescaled, ($\Tilde{X}$), all the timestep values in $\Tilde{X} = \{\Tilde{x}_1, \Tilde{x}_2, \ldots, \Tilde{x}_N\}$ are between $-1$ and $1$, i.e., $\Tilde{x}_i \in [-1, 1]$ for $i = 0, \ldots , N$. Thus, each value can be interpreted as the cosine of an imaginary angle $\varphi_i \in [0, \pi]$ \cite{wang2015imaging}:
\begin{equation}
    \Tilde{x}_i = \cos{(\varphi_i)} \quad \text{and} \quad \varphi_i = \arccos{(\Tilde{x}_i)}
    \label{eq: back_1D}
\end{equation}

\begin{figure}  [H]   
    \begin{center}
        \includegraphics[scale=0.6, trim=15mm 30mm 15mm 20mm,clip]{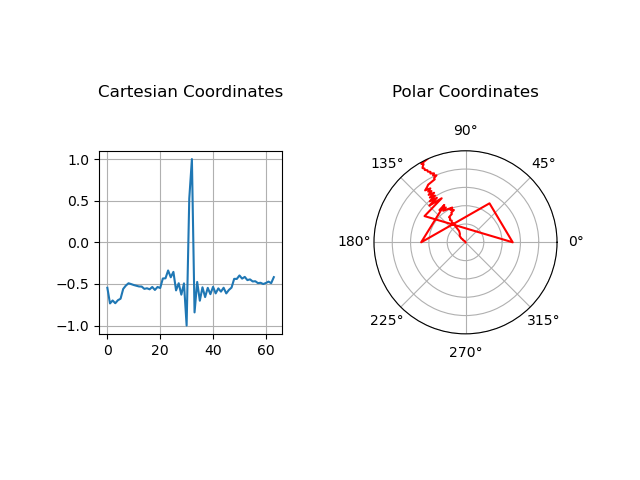}
        \caption{\emph{Normal} ECG Beat in Cartesian and Polar Coordinates}
        \label{fig: rect_polar}
    \end{center}
\end{figure}

Thus, the polar coordinates of mapped data will be (Fig. \ref{fig: rect_polar}):
\begin{equation}  
    \begin{cases}
      \varphi_i = \arccos{(\Tilde{x}_i)} & \hspace{9mm} -1\leq \Tilde{x}_i \leq1 \\
        r_i = \frac{i}{N} & \hspace{12mm}  i \in [1, N]  \\        
    \end{cases}       
\end{equation}

This mapping is \emph{bijective} (one-to-one correspondence) as $\cos{(\varphi)}$ is monotonic when $\varphi \in [0, \pi]$. Therefore, the forward as well as reverse mappings are unique. Also, the temporal relations of the timesteps are preserved in the mapping.

\subsubsection{Gramian Angular Fields (GASF/GADF)} 
After transformation into polar coordinates, the \emph{Gramian Summation Angular Field} (GASF) and the \emph{Gramian Difference Angular Field} (GADF) matrices are defined \cite{wang2015imaging}:

\begin{equation}
    GASF = 
    \begin{bmatrix}
        \cos{(\frac{\varphi_1}{2} + \frac{\varphi_1}{2})} & \ldots & \cos{(\frac{\varphi_1}{2} + \frac{\varphi_N}{2})} \\
        \vdots & \ddots & \vdots \\
        \cos{(\frac{\varphi_N}{2} + \frac{\varphi_1}{2})} & \dots & \cos{(\frac{\varphi_N}{2} + \frac{\varphi_N}{2})}
    \end{bmatrix}
    \label{eq: GASF}
\end{equation}

\vspace{2 mm}

\begin{equation}
    GADF = 
    \begin{bmatrix}
        \cos{(\frac{\varphi_1}{2} - \frac{\varphi_1}{2})} & \ldots & \cos{(\frac{\varphi_1}{2} - \frac{\varphi_N}{2})} \\
        \vdots & \ddots & \vdots \\
        \cos{(\frac{\varphi_N}{2} - \frac{\varphi_1}{2})} & \dots & \cos{(\frac{\varphi_N}{2} - \frac{\varphi_N}{2})}
    \end{bmatrix}
    \label{eq: GADF}
\end{equation}

\vspace{2 mm}

GASF and GADF are in fact quasi-Gramian matrices because the defined \emph{cos()} functions do not satisfy the linear property in the inner-product space, however, they do preserve the temporal dependency of the timesteps in the time series \cite{wang2015imaging}. Additionally, the main diagonal of GASF can be used directly in the de-embedding to reconstruct the original time series in the Cartesian coordinates by Eq. \ref{eq: back_1D}, as \emph{cos()} is monotonic when $\varphi_i \in [0, \pi]$. In GASF/GADF embedding, half-angles are used because with the full-angles, the mapping in the de-embedding will not be unique ($\cos(2\theta) = 2\cos^2(\theta) - 1 \hspace{3mm} \longrightarrow \hspace{3mm} \cos(\theta) = \pm \frac{1}{2} \sqrt{\cos(2\theta) + 1 \,}$).

\subsubsection{Markov Transition Fields (MTF)} 

GASF and GADF capture the \emph{static} information in the time series elements without any notion on the \emph{dynamic} information, i.e., how the values in timesteps change in time progression. In contrast, MTF captures the \emph{dynamic} information by setting up some $Q$ quantile bins and assigning each $x_i$ to its corresponding bin $q_i$ where $i \in [1, Q]$. 
For any pair of $x_i$ and $x_j$, with the corresponding bins $q_i$ and $q_j$, $M_{ij}$ in \emph{Matrix Transition Field} (MTF) denotes the probability of transitioning from $q_i$ to $q_j$. Thus, the MTF matrix takes into account the \emph{temporal positions} as well as \emph{temporal changes} of the timesteps in the time series. The main diagonal in the MTF matrix represents the probability of transitioning from the quantile at timestep $i$ to itself (self-transition probability) \cite{wang2015imaging}:

\begin{equation}
    GTF = 
    \begin{bmatrix}
        M_{11} & \ldots & M_{1N} \\
        \vdots & \ddots & \vdots \\
        M_{N1} & \ldots & M_{NN} \\
    \end{bmatrix}
    \label{eq: GTF}
\end{equation}

\subsection{De-embedding}
After the diffusion model is trained on the $2D$ embedded data, the trained model is sampled to generate synthetic $2D$ $3$-channel ECG data. Since the elements on the main diagonal of the GASF channel of the generated data consist only of the univariate data, $\Bar{X} = \{ \cos{(\Bar{\varphi}_1)}, \ldots, \cos{(\Bar{\varphi}_{N}}) \}$, we can de-embed the generated $2D$ data back into $1D$ space (i.e., reconstruct the time series) very easily by using Eq. \ref{eq: back_1D}, given that the mapping is bijective.

\subsection{Precision or Recall}
In this study we intend to maximize the number of true positives (TP) and at the same time minimize the number of false positives (FP). Thus, \begin{small}$\frac{FP}{TP}$\end{small} is to be minimized. Therefore, Precision score \begin{small}$(Pr=\frac{TP}{TP+FP}=\frac{1}{1+FP/TP})$\end{small} relates to the purpose of this study more than the Recall score (\begin{small}$Re=\frac{TP}{TP+FN} =\frac{1}{1+FN/TP})$\end{small}, as in the abundance of the generated data, it is immaterial if some of the \emph{good} samples are overlooked (false negatives).

\subsection{Precision - Recall Plot Area Under Curve}
One of the metrics used to evaluate the performance of binary classification is the area under the Precision-Recall plot. This plot captures the trade-off between the two scores at different (rather than at one single) thresholds. First, in the classification of the whole set of the test data, \begin{small}$Pr=\frac{TP}{TP+FP}$\end{small} is plotted against \begin{small}$Re=\frac{TP} {TP+FN}$\end{small} when the threshold of classification probability is varied from zero to one. Then the area under the curve of the plot is measured. The area is equal to $1$ for a perfect classification \cite{hastie2009elements}.

\subsection{Receiver Operator Plot Area Under Curve}
Receiver Operator Curve (ROC) is the plot of \begin{small}$TPR=\frac{TP}{TP+FN}$\end{small} against \begin{small}$FPR=\frac{FP}{FP+TN}$\end{small}. ROC AUC score is a metric which measures the trade-off between $TPR$ and $FPR$ when the threshold of the \emph{classification probability} is varied from zero to one. A higher ROC AUC value indicates a better classification model \cite{hastie2009elements}.

\subsection{Authenticity of Generated Beats}
One of the metrics we used in our comparison is the authenticity test, in which the generated beats are checked whether they can function as and replace the \emph{real} beats in a classification task. This investigation is done via the so-called \emph{classification} or \emph{authenticity} test. The objective of this binary classification test is to distinguish Normal beats from the anomaly (here we picked the typical Class L as the anomaly). 

First, the state-of-the-art classifier (ResNet34) is trained on a totally balanced and all-real training set consisting of two classes: \textit{\textbf{N}} (Normal Beat) and \textit{\textbf{L}} (Left Bundle Branch Block Beat) with $7,000$ samples in each class. The trained classifier is put to test on an unseen test set and the classification metrics are recorded. This is the \emph{reference case}, and any other study case is compared with this case (Table \ref{tab: auth test stats}). Then, the \emph{compromised case} is formed by imbalancing the training set purposely by reducing the number of samples in \textit{\textbf{N}} class down to $350$. The performance of the classification of the compromised case, which has been made poor intentionally, is recorded. Then, the imbalanced training set is augmented/balanced by the synthetic beats in each of the study cases and the classifier is trained on them. Since the training set is augmented/balanced, the classification performance improves significantly relative to the compromised case.  The case which produces the best improvements in the classification metrics has produced the most authentic synthetic beats (Table \ref{tab: auth test stats}). The same test set is used in all cases: $1000$ samples of unseen real data in each class with no synthetic beats.

\begin{table}[ht]
	\scriptsize    
 	\caption{\textit{\textbf{Authenticity Test} \\ \hspace{3mm} Training Set Supports}}       
        \label{tab: auth test stats}
 	\centering
 		{\begin{tabular}{| c | c | c | }
            
 		\hline
 			\thead{\textbf{Cases}} & \thead{\textbf{Train Set} \\ \textbf{Class N}} & \thead{\textbf{Train Set} \\ \textbf{Class L}} \\
 			\hline
 			\hline
 				\makecell{Reference (Balanced, all real)} & \makecell{7000 r} & \makecell[cc]{7000 r}  \rule[-1mm]{0pt}{4mm} \\
 			    \hline 			
     			\makecell{Compromised (Imbalanced)} & \makecell{350 r} & \makecell[cc]{7000 r} \rule[-1mm]{0pt}{4mm} \\
     			\hline     			
     			\makecell{Augmented with case 00} & \makecell{350 r + 6650 s} & \makecell[cc]{7000 r} \rule[-1mm]{0pt}{4mm} \\
     			\hline  
                    \makecell{Augmented with case 01} & \makecell{350 r + 6650 s} & \makecell[cc]{7000 r} \rule[-1mm]{0pt}{4mm} \\
     			\hline  
                    \makecell{Augmented with case 02} & \makecell{350 r + 6650 s} & \makecell[cc]{7000 r}   \rule[-1mm]{0pt}{4mm} \\
     			\hline 
        
                    \makecell{Augmented with case GAN} & \makecell{350 r + 6650 s} & \makecell[cc]{7000 r} \rule[-1mm]{0pt}{4mm} \\
     			\hline                    
            
 		\end{tabular}}   
\end{table}

\vspace{-7mm}

\begin{center}
    \begin{table}[h!]
        \scriptsize
        \begin{tabular}{llll}
           \hspace{8mm}  &  \hspace{-4mm} \emph{r: Real Beat} & \hspace{5mm} \emph{s: Synthetically Generated Beat}   & 
        \end{tabular}
    \end{table}
\end{center}

\subsubsection{Classifier: ECGResNet34}
ResNet34 \cite{he2016deep} is the state-of-the-art tool used in the classification of images. It has 34 layers and incorporates \emph{residual} building blocks. Each block is comprised of two $3\times3$ convolutional layers with a residual stream \cite{he2016deep}, which reduces the risk of gradient vanishing/exploding. It is pretrained on the ImageNet dataset (more than $100,000$ images in $200$ classes). We used its $1D$ implementation \hspace{1sp}\cite{layshukecgresnet34} for our classification test.

\section{Experimental Setup}
\subsection{WGAN-GP Model Design}
The architectures of the generator and the critic in the WGAN-GP model are comprised of building blocks that are repeated multiple times (Table \ref{tab: wgangp blocks}). The details of the architectures of the generator and the critic are shown in Table \ref{tab: archs_wgangp}.

\begin{table}[H]
	\scriptsize
 	\caption{\textit{\textbf{WGAN-GP Building Blocks}}}
 	\centering
 		\begin{tabular}{| c | c | c | }
 		\hline
 			\thead{\textbf{Layer}} & \thead{\textbf{Generator}} & \thead{\textbf{Critic}} \\
 			\hline
 			\hline 			 			
     			\makecell{1} & \makecell[cc]{ConvTranspose1d \footnote[1]{*}}  & \makecell[cc]{ConvTranspose1d \footnote[1]{*}} \rule[-2mm]{0pt}{5mm} \\
     			\hline     			
     			\makecell{2} & \makecell[cc]{BatchNorm1d}  & \makecell[cc]{InstanceNorm1d} \rule[-2mm]{0pt}{5mm} \\
     			\hline     			
     			\makecell{3} & \makecell[cc]{ReLU}  & \makecell[cc]{LeakyReLU} \rule[-2mm]{0pt}{5mm} \\
     			\hline      			 			
 		\end{tabular} 	
 	\label{tab: wgangp blocks} 
\end{table}

\vspace{-8mm}

\begin{center}
    \begin{table}[H]
        \scriptsize
        \begin{tabular}{llll}
           \hspace{10mm} 1: &  \hspace{-4mm} \emph{kernel size = 4, stride = 2, padding = 1} &   & 
        \end{tabular}
    \end{table}
\end{center}

\vspace{-5mm}

\begin{table}[H]
	\scriptsize
 	\caption{\textit{\textbf{WGAN-GP Architecture}}}

 	\centering
 		\begin{tabular}{| c | c | c | }
 		\hline
 			\thead{\textbf{Layer}} & \thead{\textbf{Generator}} & \thead{\textbf{Critic}} \\
 			\hline
 			\hline
 				\makecell{Input} & \makecell[cc]{$16\times100\times1$}  & \makecell[cc]{$16\times1\times64$} \rule[-2mm]{0pt}{5mm} \\
 			    \hline
 			
     			\makecell{1} & \makecell[cc]{Block}  & \makecell[cc]{Vonv1d, LeakyReLU} \rule[-2mm]{0pt}{5mm} \\
     			\hline
     			
     			\makecell{2} & \makecell[cc]{Block}  & \makecell[cc]{Block} \rule[-2mm]{0pt}{5mm}  \rule[-2mm]{0pt}{5mm} \\
     			\hline
     			
     			\makecell{3} & \makecell[cc]{Block}  & \makecell[cc]{Block} \rule[-2mm]{0pt}{5mm} \\
     			\hline
     			
     			\makecell{4} & \makecell[cc]{Block}  & \makecell[cc]{Block} \rule[-2mm]{0pt}{5mm} \\
     			\hline
     			
     			\makecell{5} &  \makecell[cc]{ConvTranspose1d} & \makecell[cc]{Conv1d} \rule[-2mm]{0pt}{5mm} \\
     			\hline 
     			
     			\makecell{6} &  \makecell[cc]{FC} & \makecell[cc]{FC} \rule[-2mm]{0pt}{5mm} \\
     			\hline 
     			
     			\makecell{7} &  \makecell[cc]{tanh} &  -  \rule[-2mm]{0pt}{5mm} \\ 
     			\hline
     			
     			\makecell{Output} &  \makecell[cc]{$16\times1\times64$} &  \makecell[cc]{$16\times1\times1$}  \rule[-2mm]{0pt}{5mm} \\ 
     			\hline 			
 		\end{tabular} 	
 	\label{tab: archs_wgangp} 
\end{table}

\subsection{Improved DDPM Model Design}
Improved DDPM \cite{nichol2021improved} has been used \emph{as-is} as the diffusion model. Codes of the DDPM are taken from \cite{nichol2021improved} (\url{https://github.com/openai/improved-diffusion}). 

\subsection{Platform}
The training of and sampling from the DDPMs are done on the Arc (the HPC cluster at the University of Texas at San Antonio (UTSA)). Currently, Arc can run programs with two $V$-$100$ GPUs on each node. For this study, one node from the cluster with two parallel GPUs has been used. 

For the rest of the computations, a desktop and a laptop have been used: a Dell Alienware desktop with Intel i$9$-$9900$k at $3.6$ GHz ($8$ cores, $16$ threads) microprocessor, $64$ GB RAM, and NVIDIA GeForce RTX $2080$ Ti graphics card with $24$ GB RAM, and a personal Dell G$7$ laptop with an Intel i$7$-$8750$H at $2.2$ GHz ($6$ cores, $12$ threads) microprocessor, $20$ GB of RAM, and NVIDIA GeForce $1060$ MaxQ graphics card with $6$ GB. Our codes are available on the GitHub page of the paper (\url{https://github.com/mah533/Synthetic-ECG-Signal-Generation-using-Probabilistic-Diffusion-Models}).

\subsection{Study Cases}
Three different hyperparameter settings for the Improved DDPM have been considered  which are shown in Table \ref{table: DM Cases}. The rest of the parameters are the same for all cases.

\begin{table}[ht]
	\scriptsize
 	\caption{\textit{\textbf{DM Case Studies}}}
        \label{table: DM Cases}
 	\centering
 		\begin{tabular}{| c | c | c | c | c | }
 		\hline
 			\thead{\textbf{Cases}}  & \thead{\textbf{Learn} \\ \textbf{Sigma}} & \thead{\textbf{Noise} \\ \textbf{Schedule}} & \thead{\textbf{Use KL \footnote[1]{*}}} & \thead{\textbf{Schedule} \\ \textbf{Sampler}}\\
 			\hline
 			\hline
 				\makecell{00} & \makecell[cc]{False}  & \makecell[cc]{Linear} & \makecell[cc]{False}  & \makecell[cc]{Uniform} \rule[-2mm]{0pt}{5mm} \\
 			    \hline
 			
     			\makecell{01} & \makecell[cc]{True}  & \makecell[cc]{Cosine} & \makecell[cc]{True}  & \makecell[cc]{Uniform}  \rule[-2mm]{0pt}{5mm}\\  
     			\hline
     			
     			\makecell{02} & \makecell[cc]{True}  & \makecell[cc]{Cosine} & \makecell[cc]{True}  & \makecell[cc]{loss second \\ moment}\\
     			\hline     			
 		\end{tabular} 	 	
\end{table}

\vspace{-7mm}

\begin{center}
    \begin{table}[h!]
        \scriptsize
        \begin{tabular}{llll}
           \hspace{5mm} 1: &  \hspace{-4mm} \emph{Kullback-Leibeler (KL) Divergence} &   & 
        \end{tabular}
    \end{table}
\end{center}

The above three study cases have been compared to the $4^{th}$ case, which is the synthetic ECG beats generated by the WGAN-GP. Since the ultimate goal is to generate realistic synthetic beats that resemble and function like real beats as closely as possible, an additional case (rl) is considered for reference, in which real beats have been used instead of synthetic ones in the corresponding comparison. 

 \section{Results}
\begin{figure}[htb]
    \centering
    \begin{subfigure}[b]{0.45\textwidth}
        \centering
        {\includegraphics[scale=0.25, trim=0mm 8mm 0mm 0mm,clip]{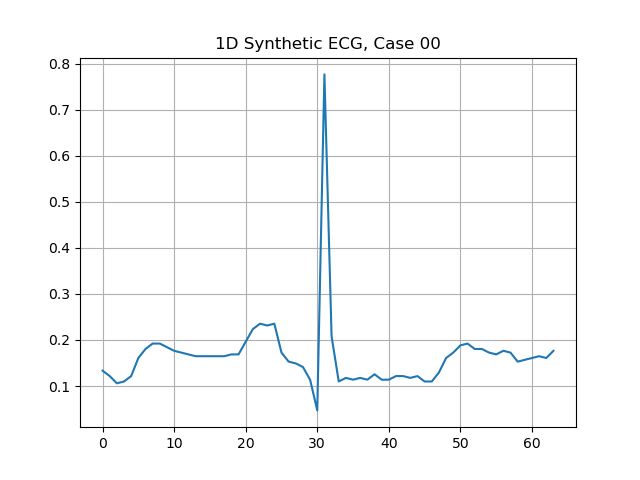}}  
        \hfill
        {\includegraphics[scale=1.5, trim=0mm 0mm 0mm 0mm,clip]{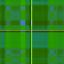}}        
        \caption{Case 00}        
    \end{subfigure}
    
    \vskip\baselineskip
    
    \begin{subfigure}[b]{0.45\textwidth}
        \centering
        {\includegraphics[scale=0.25, trim=0mm 8mm 0mm 0mm,clip]{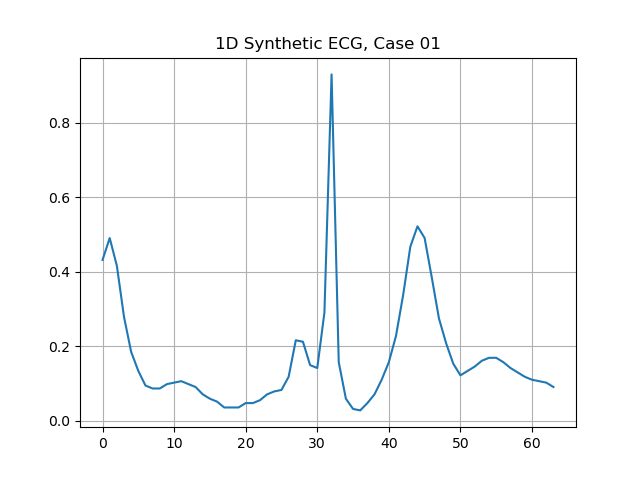}}
        \hfill
        {\includegraphics[scale=1.5, trim=0mm 0mm 0mm 0mm,clip]{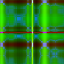}}        
        \caption{Case 01}
    \end{subfigure}

    \begin{subfigure}[b]{0.45\textwidth}
        \centering
        {\includegraphics[scale=0.25, trim=0mm 8mm 0mm 0mm,clip]{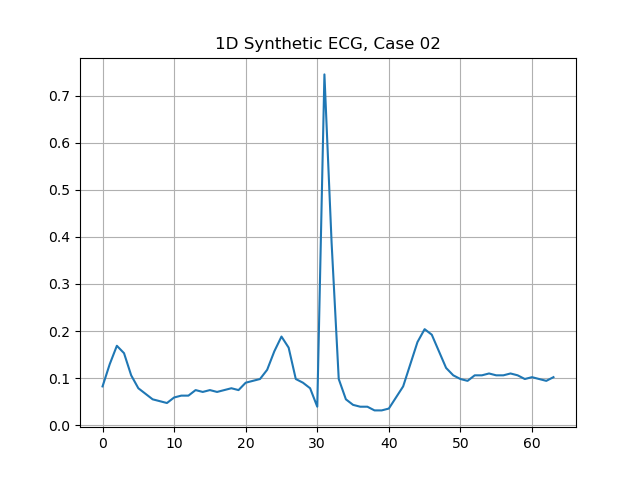}}  
        \hfill
        {\includegraphics[scale=1.5, trim=0mm 0mm 0mm 0mm,clip]{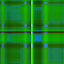}}        
        \caption{Case 02}        
    \end{subfigure}

    \begin{subfigure}[b]{0.45\textwidth}
        \centering
        {\includegraphics[scale=0.25, trim=0mm 8mm 0mm 0mm,clip]{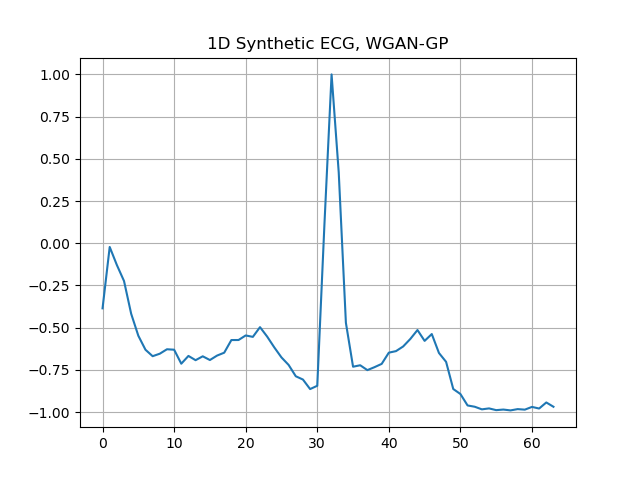}}  
        \hfill
        {\includegraphics[scale=0, trim=0mm 0mm 0mm 0mm,clip]{syn_wgangp.png}}        
        \caption{Case WGAN-GP}        
    \end{subfigure}
    
    \caption{\small Samples of Synthetically Generated ECG Signals}
    \label{fig: syn ECG all}
\end{figure}

Samples of the generated synthetic ECG signals ($1D$ and $2D$) are shown in the Figure \ref{fig: syn ECG all}.
The aforementioned four case studies are compared by the \emph{quality}, \emph{distribution} and \emph{authenticity} of the generated beats in each case.

\subsection{Quality}
When discussing the quality of the beats, we refer to how much the generated synthetic beats resemble the real ones in appearance and morphology. For assessing the quality of the heartbeats quantitatively, \emph{the average distances of the generated beats from a randomly selected template} is measured using two distance functions (AKA similarity measures): Dynamic Time Warping (DTW) and Fr\'echet distance functions. Both distance functions consistently show that the generated  beats by the WGAN-GP model are by far closer to the reference case (rl), in terms of quality (Table \ref{table: gb quality}).  

\begin{table}[ht]
	\scriptsize
 	\caption{\textit{\textbf{Quality of Generated Beats}}}
        \label{table: gb quality}
 	\centering
 		\begin{tabular}{| c | c | c | }
 		\hline
 			\thead{\textbf{Cases}} & \thead{\textbf{Ave. DTW} \\ \textbf{Distance}} & \thead{\textbf{Ave. Fr\'echet} \\ \textbf{Distance}} \\
 			\hline
 			\hline
 				\makecell{00} & \makecell[cc]{6.67}  & \makecell[cc]{1.074} \rule[-1mm]{0pt}{4mm} \\
 			    \hline 			
     			\makecell{01} & \makecell[cc]{6.95}  & \makecell[cc]{1.117} \rule[-1mm]{0pt}{4mm} \\
     			\hline     			
     			\makecell{02} & \makecell[cc]{6.36}  & \makecell[cc]{1.042} \rule[-1mm]{0pt}{4mm} \\
     			\hline  
                    \makecell{GAN} & \makecell[cc]{2.12}  & \makecell[cc]{0.723} \rule[-1mm]{0pt}{4mm} \\
     			\hline  
                    \makecell{Real (rl)} & \makecell[cc]{2.09}  & \makecell[cc]{0.718} \rule[-1mm]{0pt}{4mm} \\
     			\hline     			
 		\end{tabular} 	 	
\end{table}

\subsection{Distribution}
Maximum Mean Discrepancy (MMD) is a kernel based statistical tool to measure the distance between two \emph{distributions} (Eq. \ref{eq: MMD}) \cite{gretton2012kernel}. To compare the distributions of generated beat sets, equal number of samples (7000, i.e., the total number of generated beats) from the real data (rl) and the generated data in each case are selected randomly, ($m=n=7000$) and the MMD value between them is measured utilizing the linear kernel. For reference, the MMD value between two disjoint sets of real samples is shown as well (Table \ref{table: gb distribution}).

\scriptsize
\begin{multline}
\label{eq: MMD}
    MMD^{2} \left(p,q\right) = 
    \\ 
    \mathbb{E}_{x,x'} \left[k\left(x_i,x'_{j}\right)\right] - 2\mathbb{E}_{x,y} \left[k\left(x_i,y_{j}\right)\right] + \mathbb{E}_{y,y'} 
    \left[k\left(y_i,y'_{j}\right)\right] 
    \\
     = \frac{1}{m(m-1)} \sum_{i=1}^{m} \sum_{j\neq i}^{m} k\left(x_i,x_j\right) -\frac{2}{mn} \sum_{i=1}^{m} \sum_{j=1}^{n} k\left(x_i,x_j\right) \\ 
    + \frac{1}{n(n-1)} \sum_{i=1}^{n} \sum_{j\neq i}^{n} k\left(y_i,y_j\right)  \hspace{30mm}\\
\end{multline}

\begin{table} [H]    
    \caption{\textit{\textbf{MMD Value of Synthetic and Real Beats}}}
    \label{table: gb distribution}
    \centering
    \scriptsize
    \begin{tabular}{|P{8mm}|P{9mm}|P{8mm}| P{8mm}|P{10mm}| P{8mm}|}
        \hline
        \thead{\textbf{Cases}} & \thead{\textbf{00-rl}} & \thead{\textbf{01-rl}} & \thead{\textbf{02-rl}} & \thead{\textbf{GAN-rl}} & \thead{\textbf{rl-rl}}\\
        
        \hline
        \hline
         \textbf{MMD} & 39.8  &  44  & 35.9  &  1.00 & 0.0 \rule[-1mm]{0pt}{4mm} \\
        \hline        
        \end{tabular}        
\end{table}

\normalsize
In terms of the distribution of the generated beats, the WGAN-GP model generates beats much closer to the real beats than the diffusion models do.

\subsection{Authenticity}
Here we quantitatively measure how much the generated beats can replace (i.e., function as) the real ones in a classification test. The metrics used for the authenticity tests are $\bm{(A)}$ Average Precision scores, $\bm{(B)}$ the Area Under the Curve (AUC) of the Precision-Recall Curves, as well as $\bm{(C)}$ the AUC of the Receiver Operating Characteristic curves (AUC ROC score).

\begin{table}[ht]
	\scriptsize
 	\caption{\textit{\textbf{Authenticity of Generated Beats}}}
        \label{table: gb auth test}
 	\centering
 		\begin{tabular}{| c | c | c | c |}
 		\hline
 			\thead{\textbf{Cases}} & \thead{\textbf{Ave} \\ \textbf{Precision}} & \thead{\textbf{PRC \footnote[1]{*}} \\ \textbf{AUC \footnote[2]{*} Score}} & \thead{\textbf{ROC \footnote[3]{*}} \\ \textbf{AUC Score}} \\
 			\hline
 			\hline
 				\makecell{00} & \makecell[cc]{0.90}  & \makecell[cc]{0.95}  & \makecell[cc]{0.96}  \rule[-1mm]{0pt}{4mm} \\
 			    \hline 			
     			\makecell{01} & \makecell[cc]{0.55}  & \makecell[cc]{0.68} & \makecell[cc]{0.63} \rule[-1mm]{0pt}{4mm} \rule[-1mm]{0pt}{4mm} \\
     			\hline     			
     			\makecell{02} & \makecell[cc]{0.76}  & \makecell[cc]{0.76}  & \makecell[cc]{0.81} \rule[-1mm]{0pt}{4mm} \\
     			\hline  
                    \makecell{GAN} & \makecell[cc]{0.96}  & \makecell[cc]{0.99} & \makecell[cc]{0.99} \rule[-1mm]{0pt}{4mm} \\
     			\hline  
                    \makecell{Real (rl)} & \makecell[cc]{0.98}  & \makecell[cc]{1.00} & \makecell[cc]{1.00}  \rule[-1mm]{0pt}{4mm} \\
     			\hline     			
 		\end{tabular} 	 	
\end{table}

\vspace{-7mm}

\begin{center}
    \begin{table}[h!]
        \scriptsize
        \begin{tabular}{llll}
           \hspace{7mm} 1: &  \hspace{-4mm} \emph{Precision-Recall Curve} &   &  \\
           \hspace{7mm} 2: &  \hspace{-4mm} \emph{Area Under Curve} &   &  \\
           \hspace{7mm} 3: &  \hspace{-4mm} \emph{Receiver Operating Characteristic Curve} &   &  
        \end{tabular}
    \end{table}
\end{center}

The Average Precision scores show that WGAN-GP model outperforms the DDPM in correctly classifying the beats, minimizing FP and maximizing TP. It should be noted that micro- and macro-averages are the same as the test set is balanced (Table \ref{table: gb auth test}). 

 \subsubsection{Precision - Recall Curves}
\begin{figure} [H]  
    \centering
    \includegraphics[scale=0.4, trim=5mm 5mm 5mm 5mm,clip]{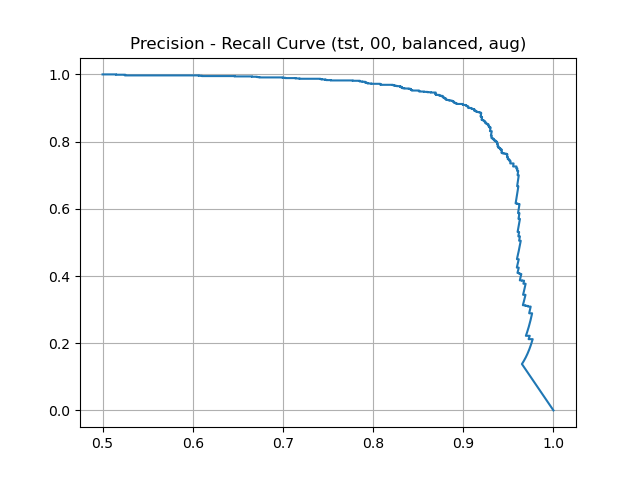}
    \caption{Precision-Recall Curve, Case 00}
    \label{fig: pr curve 00}    
\end{figure}

\begin{figure} [H]   
    \centering
    \includegraphics[scale=0.4, trim=5mm 5mm 5mm 5mm,clip]{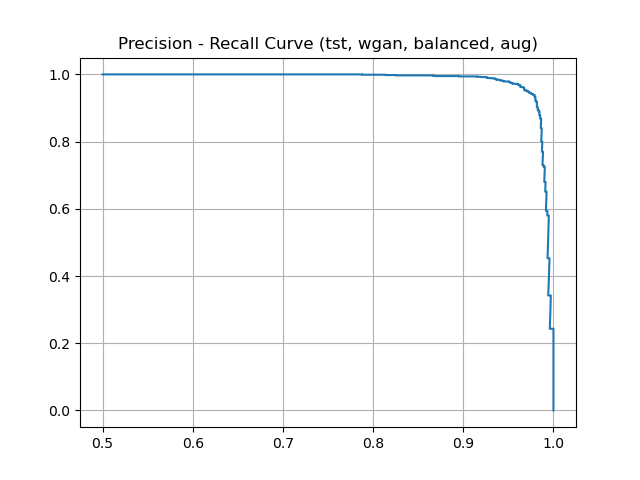}
    \caption{Precision-Recall Curve, Case GAN}
    \label{fig: pr curve wgan}    
\end{figure}

\begin{figure} [H]   
    \begin{center}
        \includegraphics[scale=0.4, trim=5mm 5mm 5mm 5mm,clip]{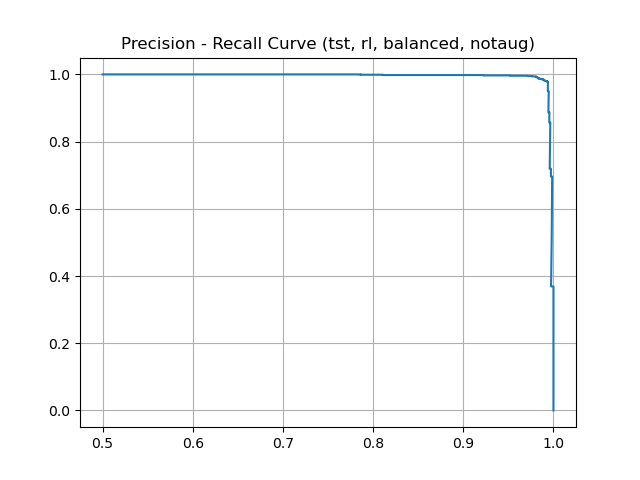}
        \caption{Precision-Recall Curve, Case real}
        \label{fig: pr curve rl}
    \end{center}
\end{figure}

 The Precision-Recall curves' Area Under Curve, (PR AUC), score is a better metric than the Average Precision score, which is calculated at only one threshold. By checking the Precision-Recall curves visually, it can be seen that the WGAN-GP model (Figure \ref{fig: pr curve wgan}) produces a graph very close to the real case (Figure \ref{fig: pr curve rl}). However, case $00$ has the best PR curve among the DDPM cases (Figure \ref{fig: pr curve 00}). Also, the Precision-Recall AUC score (Table \ref{table: gb auth test}) confirms the visual check. The same argument holds for the ROC AUC score. 

\begin{figure} [H]    
    \centering
    \includegraphics[scale=0.4, trim=5mm 5mm 5mm 5mm,clip]{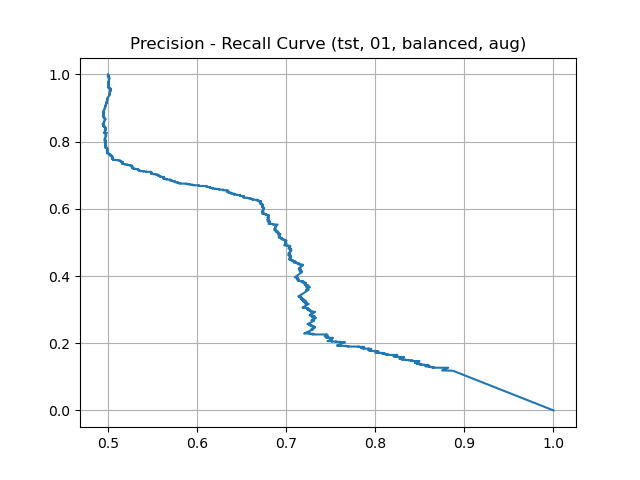}
    \caption{Precision-Recall Curve, Case 01}
    \label{fig: pr curve 01}    
\end{figure}

\begin{figure} [H]  
    \centering
    \includegraphics[scale=0.4, trim=5mm 5mm 5mm 5mm,clip]{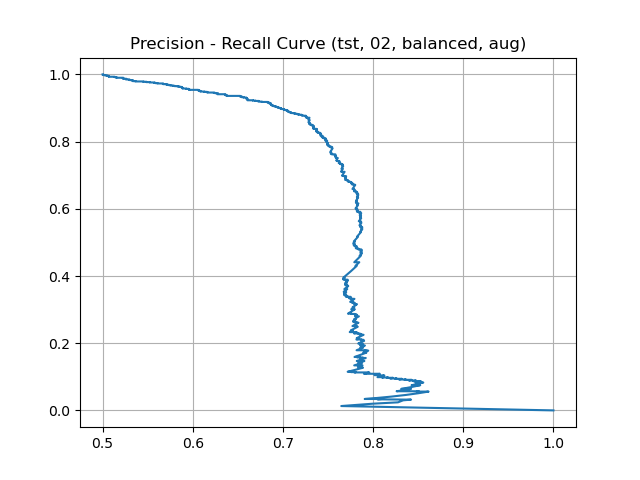}
    \caption{Precision-Recall Curve, Case 02}
    \label{fig: pr curve 02}    
\end{figure}

\subsubsection{Confusion Matrix}
The elements on the main diagonal of the confusion matrix show the percentage of number of times the beats are classified correctly. Again, the synthetic beats generated by the WGAN-GP model behave much more like real beats than the beats generated by DDPM do. 
\begin{table} [H]
 \scriptsize
\caption{\textit{\textbf{Confusion Matrices (all values are in \%)}}}
  \label{tab: cm all}
  
    \begin{tabular}{c c}
        \begin{subtable} [t] {.45\linewidth}
        \centering
        
             \begin{tabular}{|P{8mm}|P{8mm}|P{8mm}|}
                \hline
                 & {\textit{\textbf{Pred. N}}}  &  {\textit{\textbf{Pred. L}}}  \\
                \hline
               {\textit{\textbf{Real N}}} & 85.7  &  14.3   \\
                \hline
                 {\textit{\textbf{Real L}}} & 5.4 & 94.6     \\       
                \hline        
            \end{tabular}        
        \caption{\scriptsize Case 00}
        \label{tab: cm 00}
      \end{subtable}%
      
       & 
       
       \begin{subtable} [t] {.45\linewidth}
        \centering
       \begin{tabular}{|P{8mm}|P{8mm}|P{8mm}|}
           \hline
            & {\textit{\textbf{Pred. N}}}  &  {\textit{\textbf{Pred. L}}}  \\
            \hline
           {\textit{\textbf{Real N}}} & 37.1  &  62.9    \\
            \hline
            {\textit{\textbf{Real L}}}&  28.2 & 71.8      \\       
            \hline        
        \end{tabular}        
        \caption{\scriptsize Case 01}
        \label{tab: cm 01}
      \end{subtable}
    
    \\
    
    \begin{subtable} [t] {.45\linewidth}
        \centering
       \begin{tabular}{|P{8mm}|P{8mm}|P{8mm}|}
            \hline
            & {\textit{\textbf{Pred. N}}}  &  {\textit{\textbf{Pred. L}}}  \\
            \hline
          {\textit{\textbf{Real N}}} &  73.6  &  26.4    \\
            \hline
           {\textit{\textbf{Real L}}} & 21.0 & 79.0      \\       
            \hline        
        \end{tabular}        
        \caption{\scriptsize Case 02}
        \label{tab: cm 02}
      \end{subtable}
      
  &

   \begin{subtable} [t] {.45\linewidth}
        \centering
       \begin{tabular}{|P{8mm}|P{8mm}|P{8mm}|}
            \hline
                & {\textit{\textbf{Pred. N}}}  &  {\textit{\textbf{Pred. L}}}  \\
            \hline
         {\textit{\textbf{Real N}}}  & 93.9  &  6.1    \\
            \hline
           {\textit{\textbf{Real L}}} & 1.8 & 98.2      \\       
            \hline        
        \end{tabular}        
        \caption{\scriptsize Case GAN}
        \label{tab: cm 03}
      \end{subtable}

    \\

        \begin{subtable} [t] {.45\linewidth}
        \centering
       \begin{tabular}{|P{8mm}|P{8mm}|P{8mm}|}
            \hline
                & {\textit{\textbf{Pred. N}}}  &  {\textit{\textbf{Pred. L}}}  \\
            \hline
         {\textit{\textbf{Real N}}}  & 99.4  &  0.6    \\
            \hline
          {\textit{\textbf{Real L}}}  & 3.4 & 96.6      \\       
            \hline        
        \end{tabular}        
        \caption{\scriptsize Case real}
        \label{tab: cm rl}
      \end{subtable}
      
    \end{tabular}  
  \end{table}

\section{Discussion}
In a typical classification task, since there is no need to map the data back into their original space, no \emph{invertibility} condition is required for the embedding. As long as there is one \emph{and only one} corresponding element in the embedded space, and no two datapoints in the original space have the same mapped datapoint in the destination space (i.e., \emph{injective} mapping), mapping is acceptable. However, in generation tasks, the generated data must be de-embedded and mapped back into the original space. Therefore, the mapping must be \emph{bijective}. That is why the spectrograms are useful in classification tasks but not in generation tasks.

Each ECG beat in our dataset has 64 timesteps. Therefore with the WGAN-GP model, each time 64 pieces of information are generated per heartbeat, $100\%$ of which are \emph{useful information}, from the user's perspective. Whereas in DDPM, $3\times64\times64=12,288$ pieces of information are generated per beat, from which only 64 pieces ($5.2\%$) are useful/extracted and the rest are discarded. In the training of the deep learning models, the optimization process finds the optimum values of the trainable parameters in a way that each piece of the generated information are in some certain neighborhood. The total error in the loss function is \emph{typically} comprised of $64$ elements in the WGAN-GP and $12,288$ elements in DDPM per beat. Thus, a tighter neighborhood in WGAN-GP (closer to the real beats) and a more relaxed one in DDPM is quite natural.

\section{Conclusion}
\subsection{what we did}
In this paper, we presented a pipeline to generate synthetic $1D$ ECG time series using the $2D$ probabilistic diffusion model (Improved DDPM). 

\subsection{Why we did it}
With the remarkable success of $2D$ computer vision models, specifically the Improved DDPM and its superiority to the GAN models \cite{nichol2021improved} \cite{dhariwal2021diffusion}) and the widespread availability of their pretrained version, it makes sense to apply them to $1D$ time series. One of the benefits of the processing of the data in $2D$ space is providing additional data augmentation techniques (such as flipping, rotation, and mirroring), which are helpful specifically in classification tasks.

\subsection{How we did it}
In this study, we used unconditional models and used only the $N$ class (\emph{Normal Sinus Beat}) from the MIT-BIH Arrhythmia dataset \cite{moody2001impact} \cite{goldberger2000components}. The general pipeline used in this study is shown in Figure. \ref{fig: pipeline}. First, the $1D$ ECG time series are transformed into the polar coordinates. Then, they are embedded into the $3$-channel $2D$ space, similar to RGB image files. \emph{Gramian Summation/Difference Fields} (GASF/GADF) and \emph{Markov Transition Field} (MTF) are used to produce the three $2D$ matrices, which are then put together to form one single image file for each beat. The Improved DDPM model \cite{nichol2021improved} is trained and sampled  to generate $2D$ ECG signals, which are then de-embedded to reconstruct a $1D$ ECG signal.
The generated data by DDPM are in $3$ study cases with $3$ different settings in hyperparameters. They are compared to the synthetically generated data by WGAN-GP, as well as to the \emph{real beats}, to see how close the cases are to reality, since the ultimate goal is to generate realistic synthetic ECG beats.
The comparison is done in terms of the \emph{quality}, \emph{distribution}, and \emph{authenticity} of the data (i.e., up to what extent they can replace the \emph{real} beats in data augmentation for classification tasks).

\subsection{results summary}
The results show that the synthetic beats generated by the WGAN-GP model are consistently closer to the real beats than the beats from DDPM, in all cases and by all the metrics used. The average distances of the beats from a template (measured by the DTW and Fr\'echet Distance functions) as well as the distribution of generated beats (measured by MMD)  respectively reveal that the \emph{quality} and the \emph{distribution} of the beats from the WGAN-GP are much closer to the corresponding real ones than those of the DDPM. For quantifying the \emph{authenticity} of the generated beats, we used a classification test using the-state-of-the-art classifier ResNet34 \cite{he2016deep} and several standard classification metrics, namely average precision score, AUC of Precision-Recall curves, and AUC of ROC curves, all of which constantly show that the WGAN-GP model outperforms the Improved DDPM in this regard. 

\section{Limitations and Future Works}

The Improved DDPM \cite{nichol2021improved} developed by OpenAI is a $2D$ model, i.e., the input/output data and the processing, are in $2D$ space and it has been used in this study \emph{as-is} with almost no changes. We proposed the pipeline in Figure \ref{fig: pipeline} for this application, where the $1D$ ECG data are mapped into a $2D$ space, converted into image files, and fed to a DDPM model. The processing takes place in the $2D$ space and the generated $2D$ data is de-embedded back into $1D$ space, where the $1D$ ECG data are reconstructed, whereas the WGAN-GP model developed and used in this study is inherently $1D$, i.e., the input/output data and the processing are all in $1D$ space, with no embedding necessary. It should be emphasized that the conclusions drawn here apply only to the setting and the pipeline used in this research, and the results might be different in any other setting.

Although the probabilistic diffusion model is being applied to images and $2D$ data mostly, the general concept can be applied to $1D$ data as well, i.e., the model can take $1D$ data, perform the noising/denoising processes in $1D$ space, and generate synthetic data in $1D$ without any embedding. In this case, the model would be a better representative of the diffusion concept and the comparison would be more realistic.


\bibliographystyle{IEEEtran}
\bibliography{references}

\begin{thebibliography}{10}
\providecommand{\url}[1]{#1}
\csname url@samestyle\endcsname
\providecommand{\newblock}{\relax}
\providecommand{\bibinfo}[2]{#2}
\providecommand{\BIBentrySTDinterwordspacing}{\spaceskip=0pt\relax}
\providecommand{\BIBentryALTinterwordstretchfactor}{4}
\providecommand{\BIBentryALTinterwordspacing}{\spaceskip=\fontdimen2\font plus
\BIBentryALTinterwordstretchfactor\fontdimen3\font minus
  \fontdimen4\font\relax}
\providecommand{\BIBforeignlanguage}[2]{{%
\expandafter\ifx\csname l@#1\endcsname\relax
\typeout{** WARNING: IEEEtran.bst: No hyphenation pattern has been}%
\typeout{** loaded for the language `#1'. Using the pattern for}%
\typeout{** the default language instead.}%
\else
\language=\csname l@#1\endcsname
\fi
#2}}
\providecommand{\BIBdecl}{\relax}
\BIBdecl

\bibitem{saadatnejad2019lstm}
S.~Saadatnejad, M.~Oveisi, and M.~Hashemi, ``Lstm-based ecg classification for
  continuous monitoring on personal wearable devices,'' \emph{IEEE journal of
  biomedical and health informatics}, vol.~24, no.~2, pp. 515--523, 2019.

\bibitem{chawla2002smote}
N.~V. Chawla, K.~W. Bowyer, L.~O. Hall, and W.~P. Kegelmeyer, ``Smote:
  synthetic minority over-sampling technique,'' \emph{Journal of artificial
  intelligence research}, vol.~16, pp. 321--357, 2002.

\bibitem{lin2017focal}
T.-Y. Lin, P.~Goyal, R.~Girshick, K.~He, and P.~Doll{\'a}r, ``Focal loss for
  dense object detection,'' in \emph{Proceedings of the IEEE international
  conference on computer vision}, 2017, pp. 2980--2988.

\bibitem{wang2020generalizing}
Y.~Wang, Q.~Yao, J.~T. Kwok, and L.~M. Ni, ``Generalizing from a few examples:
  A survey on few-shot learning,'' \emph{ACM Computing Surveys (CSUR)},
  vol.~53, no.~3, pp. 1--34, 2020.

\bibitem{goodfellow2014generative}
I.~Goodfellow, J.~Pouget-Abadie, M.~Mirza, B.~Xu, D.~Warde-Farley, S.~Ozair,
  A.~Courville, and Y.~Bengio, ``Generative adversarial nets,'' \emph{Advances
  in neural information processing systems}, vol.~27, pp. 2672--2680, 2014.

\bibitem{doersch2016tutorial}
C.~Doersch, ``Tutorial on variational autoencoders,'' \emph{arXiv preprint
  arXiv:1606.05908}, 2016.

\bibitem{frid2018synthetic}
M.~Frid-Adar, E.~Klang, M.~Amitai, J.~Goldberger, and H.~Greenspan, ``Synthetic
  data augmentation using gan for improved liver lesion classification,'' in
  \emph{2018 IEEE 15th international symposium on biomedical imaging (ISBI
  2018)}.\hskip 1em plus 0.5em minus 0.4em\relax IEEE, 2018, pp. 289--293.

\bibitem{kuznetsov2020electrocardiogram}
V.~Kuznetsov, V.~Moskalenko, and N.~Y. Zolotykh, ``Electrocardiogram generation
  and feature extraction using a variational autoencoder,'' \emph{arXiv
  preprint arXiv:2002.00254}, 2020.

\bibitem{jun2018ecg}
T.~J. Jun, H.~M. Nguyen, D.~Kang, D.~Kim, D.~Kim, and Y.-H. Kim, ``Ecg
  arrhythmia classification using a 2-d convolutional neural network,''
  \emph{arXiv preprint arXiv:1804.06812}, 2018.

\bibitem{nichol2021improved}
A.~Q. Nichol and P.~Dhariwal, ``Improved denoising diffusion probabilistic
  models,'' in \emph{International Conference on Machine Learning}.\hskip 1em
  plus 0.5em minus 0.4em\relax PMLR, 2021, pp. 8162--8171.

\bibitem{dhariwal2021diffusion}
P.~Dhariwal and A.~Nichol, ``Diffusion models beat gans on image synthesis,''
  \emph{Advances in Neural Information Processing Systems}, vol.~34, pp.
  8780--8794, 2021.

\bibitem{wang2015imaging}
Z.~Wang and T.~Oates, ``Imaging time-series to improve classification and
  imputation,'' in \emph{Twenty-Fourth International Joint Conference on
  Artificial Intelligence}, 2015.

\bibitem{moody2001impact}
G.~B. Moody and R.~G. Mark, ``The impact of the {MIT-BIH} arrhythmia
  database,'' \emph{IEEE Engineering in Medicine and Biology Magazine},
  vol.~20, no.~3, pp. 45--50, 2001.

\bibitem{goldberger2000components}
A.~Goldberger, L.~Amaral, L.~Glass, J.~Hausdorff, P.~C. Ivanov, R.~Mark,
  J.~Mietus, G.~Moody, C.~Peng, and H.~Stanley, ``Components of a new research
  resource for complex physiologic signals,'' \emph{PhysioBank, PhysioToolkit,
  and Physionet}, 2000.

\bibitem{alcaraz2023diffusion}
J.~M.~L. Alcaraz and N.~Strodthoff, ``Diffusion-based conditional ecg
  generation with structured state space models,'' \emph{arXiv preprint
  arXiv:2301.08227}, 2023.

\bibitem{wang2019ecg}
P.~Wang, B.~Hou, S.~Shao, and R.~Yan, ``{ECG} arrhythmias detection using
  auxiliary classifier generative adversarial network and residual network,''
  \emph{Ieee Access}, vol.~7, pp. 100\,910--100\,922, 2019.

\bibitem{delaney1909synthesis}
A.~Delaney, E.~Brophy, and T.~Ward, ``Synthesis of realistic {ECG} using
  generative adversarial networks. arxiv 2019,'' \emph{arXiv preprint
  arXiv:1909.09150}, 2019.

\bibitem{esteban2017real}
C.~Esteban, S.~L. Hyland, and G.~R{\"a}tsch, ``Real-valued (medical) time
  series generation with recurrent conditional gans,'' \emph{arXiv preprint
  arXiv:1706.02633}, 2017.

\bibitem{adib2021synthetic}
E.~Adib, F.~Afghah, and J.~J. Prevost, ``Synthetic {ECG} signal generation
  using generative neural networks,'' \emph{arXiv preprint arXiv:2112.03268},
  2021.

\bibitem{adib2022arrhythmia}
------, ``Arrhythmia classification using cgan-augmented {ECG} signals,''
  \emph{arXiv preprint arXiv:2202.00569}, 2022.

\bibitem{ahmad2021ecg}
Z.~Ahmad, A.~Tabassum, L.~Guan, and N.~M. Khan, ``Ecg heartbeat classification
  using multimodal fusion,'' \emph{IEEE Access}, vol.~9, pp.
  100\,615--100\,626, 2021.

\bibitem{cai2022electrocardiogram}
H.~Cai, L.~Xu, J.~Xu, Z.~Xiong, and C.~Zhu, ``Electrocardiogram signal
  classification based on mix time-series imaging,'' \emph{Electronics},
  vol.~11, no.~13, p. 1991, 2022.

\bibitem{diker2019novel}
A.~Diker, Z.~C{\"o}mert, E.~Avc{\i}, M.~To{\u{g}}a{\c{c}}ar, and B.~Ergen, ``A
  novel application based on spectrogram and convolutional neural network for
  ecg classification,'' in \emph{2019 1st International Informatics and
  Software Engineering Conference (UBMYK)}.\hskip 1em plus 0.5em minus
  0.4em\relax IEEE, 2019, pp. 1--6.

\bibitem{izci2019cardiac}
E.~Izci, M.~A. Ozdemir, M.~Degirmenci, and A.~Akan, ``Cardiac arrhythmia
  detection from 2d ecg images by using deep learning technique,'' in
  \emph{2019 Medical Technologies Congress (TIPTEKNO)}.\hskip 1em plus 0.5em
  minus 0.4em\relax IEEE, 2019, pp. 1--4.

\bibitem{hao2019spectro}
C.~Hao, S.~Wibowo, M.~Majmudar, and K.~S. Rajput, ``Spectro-temporal feature
  based multi-channel convolutional neural network for ecg beat
  classification,'' in \emph{2019 41st Annual International Conference of the
  IEEE Engineering in Medicine and Biology Society (EMBC)}.\hskip 1em plus
  0.5em minus 0.4em\relax IEEE, 2019, pp. 5642--5645.

\bibitem{huang2019ecg}
J.~Huang, B.~Chen, B.~Yao, and W.~He, ``Ecg arrhythmia classification using
  stft-based spectrogram and convolutional neural network,'' \emph{IEEE
  access}, vol.~7, pp. 92\,871--92\,880, 2019.

\bibitem{oliveira2019novel}
A.~T. Oliveira and E.~G. Nobrega, ``A novel arrhythmia classification method
  based on convolutional neural networks interpretation of electrocardiogram
  images,'' in \emph{2019 IEEE International Conference on Industrial
  Technology (ICIT)}.\hskip 1em plus 0.5em minus 0.4em\relax IEEE, 2019, pp.
  841--846.

\bibitem{salem2018ecg}
M.~Salem, S.~Taheri, and J.-S. Yuan, ``Ecg arrhythmia classification using
  transfer learning from 2-dimensional deep cnn features,'' in \emph{2018 IEEE
  biomedical circuits and systems conference (BioCAS)}.\hskip 1em plus 0.5em
  minus 0.4em\relax IEEE, 2018, pp. 1--4.

\bibitem{mathunjwa2021ecg}
B.~M. Mathunjwa, Y.-T. Lin, C.-H. Lin, M.~F. Abbod, and J.-S. Shieh, ``Ecg
  arrhythmia classification by using a recurrence plot and convolutional neural
  network,'' \emph{Biomedical Signal Processing and Control}, vol.~64, p.
  102262, 2021.

\bibitem{mathunjwa2022ecg}
B.~M. Mathunjwa, Y.-T. Lin, C.-H. Lin, M.~F. Abbod, M.~Sadrawi, and J.-S.
  Shieh, ``Ecg recurrence plot-based arrhythmia classification using
  two-dimensional deep residual cnn features,'' \emph{Sensors}, vol.~22, no.~4,
  p. 1660, 2022.

\bibitem{zhang2021recurrence}
H.~Zhang, C.~Liu, Z.~Zhang, Y.~Xing, X.~Liu, R.~Dong, Y.~He, L.~Xia, and
  F.~Liu, ``Recurrence plot-based approach for cardiac arrhythmia
  classification using inception-resnet-v2,'' \emph{Frontiers in Physiology},
  p. 558, 2021.

\bibitem{kingma2019introduction}
D.~P. Kingma and M.~Welling, ``An introduction to variational autoencoders,''
  \emph{arXiv preprint arXiv:1906.02691}, 2019.

\bibitem{ho2020denoising}
J.~Ho, A.~Jain, and P.~Abbeel, ``Denoising diffusion probabilistic models,''
  \emph{Advances in Neural Information Processing Systems}, vol.~33, pp.
  6840--6851, 2020.

\bibitem{razavi2019generating}
A.~Razavi, A.~Van~den Oord, and O.~Vinyals, ``Generating diverse high-fidelity
  images with vq-vae-2,'' \emph{Advances in neural information processing
  systems}, vol.~32, 2019.

\bibitem{henighan2020scaling}
T.~Henighan, J.~Kaplan, M.~Katz, M.~Chen, C.~Hesse, J.~Jackson, H.~Jun, T.~B.
  Brown, P.~Dhariwal, S.~Gray \emph{et~al.}, ``Scaling laws for autoregressive
  generative modeling,'' \emph{arXiv preprint arXiv:2010.14701}, 2020.

\bibitem{hastie2009elements}
T.~Hastie, R.~Tibshirani, J.~H. Friedman, and J.~H. Friedman, \emph{The
  elements of statistical learning: data mining, inference, and
  prediction}.\hskip 1em plus 0.5em minus 0.4em\relax Springer, 2009, vol.~2.

\bibitem{he2016deep}
K.~He, X.~Zhang, S.~Ren, and J.~Sun, ``Deep residual learning for image
  recognition,'' in \emph{Proceedings of the IEEE conference on computer vision
  and pattern recognition}, 2016, pp. 770--778.

\bibitem{layshukecgresnet34}
A.~Lyashuk, ``{ECG} {C}lassification,''
  https://github.com/lxdv/ecg-classification/blob/master/README.md, 2021.

\bibitem{gretton2012kernel}
A.~Gretton, K.~M. Borgwardt, M.~J. Rasch, B.~Sch{\"o}lkopf, and A.~Smola, ``A
  kernel two-sample test,'' \emph{The Journal of Machine Learning Research},
  vol.~13, no.~1, pp. 723--773, 2012.

\bibitem{arjovsky2017wasserstein}
M.~Arjovsky, S.~Chintala, and L.~Bottou, ``Wasserstein {GAN},'' \emph{arXiv
  preprint arXiv:1701.07875}, 2017.

\bibitem{gulrajani2017improved}
I.~Gulrajani, F.~Ahmed, M.~Arjovsky, V.~Dumoulin, and A.~C. Courville,
  ``Improved training of wasserstein {GAN}s,'' in \emph{Advances in neural
  information processing systems}, 2017, pp. 5767--5777.

\end{thebibliography}

\newpage

\begin{appendices}
    \setcounter{equation}{0}
    \renewcommand\theequation{A.\arabic{equation}}

    \section{GAN, WGAN and WGAN-GP Models}
    \label{app: B GAN}
    Generative Adversarial Networks are \emph{two-player zero-sum minimax} games with the following loss function:
    \begin{multline}
        \min_{G}\max_{D} \mathbb{E}_{x\sim p_{\text{data}}(x)}\left[ \log{D(x)} \right] + \\ \mathbb{E}_{z\sim p_{\text{z}}(z)}\left[1 - \log{D(G(z))} \right] 
        \label{eq: appB 1}
    \end{multline}
    
    Unlike variational method models, in which the objective is to find the true distribution of the data \cite{kingma2019introduction}, in GAN models the constraints are on the expectation of the outputs of the generator and the discriminator. Therefore, it is quite possible that the GAN model focuses only on a few modality of the distribution and still the loss function is satisfied (mode collapse). Moreover, the parameters oscillate and convergence is not achieved always. 
    
    Using the the Wasserstein distance function (Kontorovich-Rubenstein Duality) \cite{arjovsky2017wasserstein} improves the convergence: 
    
    \begin{equation}
         W_1\left(p_r, q_g \right) =  \sup_{{\lVert f \rVert}_L \le1} \mathop{{}\mathbb{E}_{x\sim p_r}} \left[f(x)\right] - \mathop{{}\mathbb{E}_{x\sim p_g}} \left[f(x)\right] 
         \label{eq: appB 2}
    \end{equation}
    
    However, the constraint $\lVert f \rVert_L \le 1$  requires $f(.)$ be $1$-Lipschitz. Then the loss function of WGAN becomes:
    \begin{multline}
        \min_{G}\max_{D} \mathbb{E}_{x\sim p_{\text{data}}(x)}\left[ {D(x)} \right] - \mathbb{E}_{x\sim p_{\text{g}}(\Tilde{x})}\left[{D(\Tilde{x})} \right]       
        \label{eq: appB 3}
    \end{multline}
    
    \emph{Parameter clipping} can be used to assure the Lipschitz condition, i.e., keeping the magnitude of the parameters bounded $\left[ -c, c \right]$, which can easily lead to optimization difficulties (vanishing/exploding gradient) \cite{gulrajani2017improved}. 
    
    Gulrajani \emph{et al.} \cite{gulrajani2017improved} proposed the WGAN with Gradient Penalty (WGAN-GP) in which, instead of parameter clipping, they penalized (regularize) the loss function with the magnitude of the gradient as a constraint, using a Lagrange multiplier $\lambda$:
    
    \begin{multline}
        \min_{G}\max_{D} \mathbb{E}_{x\sim p_{\text{data}}(x)}\left[ {D(x)} \right] - \mathbb{E}_{x\sim p_{\text{g}}(\Tilde{x})}\left[{D(\Tilde{x})} \right]  +\\ 
        \lambda \enspace \mathbb{E}_{\hat{x} \sim p_{\text{data}}(x)}\left[ {\left( {\lVert \nabla_{\hat{x}} D \left( \hat{x} \right) \rVert}_2 - 1 \right)}^2  \right]    
        \label{eq: appB 4}
    \end{multline}
\end{appendices}

\end{document}